\documentclass[final]{svjour2}
\usepackage{graphicx}
\usepackage{rotating}
\usepackage{amssymb}
\usepackage{mathptmx}
\usepackage[numbers]{natbib}
\usepackage{bm}
\usepackage{pstricks}

\newcommand{\eqref}[1]{(\ref{#1})}

\makeatletter
\journalname{Journal of Low Temperature Physics}

\bibpunct{}{}{,}{s}{}{,}

\begin{document}

\newcommand{\hdblarrow}{H\makebox[0.9ex][l]{$\downdownarrows$}-}
\title{Detection of entanglement
\\
 in ultracold lattice gases}

\author{G. De Chiara$^1$ \and A. Sanpera$^{2,1}$}

\institute{1:F\'isica Te\`orica: Informaci\'o i Processos Qu\`antics, Universitat Aut\`{o}noma de Barcelona, E-08193 Bellaterra, Spain
\\2:ICREA, Instituci\`o Catalana de Recerca i Estudis Avan\c{c}ats, E08011 Barcelona }

\date{\today}

\maketitle


\begin{abstract}
We propose the use of quantum polarization spectroscopy for detecting multi-particle entanglement of ultracold atoms in optical lattices. This method, based on  a light-matter interface employing the quantum Farady effect, allows for the non destructive measurement of spin-spin correlations. We apply it to the specific example of a one dimensional spin chain and reconstruct its phase diagram using the light signal, readily measurable in experiments with ultracold atoms. Interestingly, the same technique can be extended to detect quantum many-body entanglement in such systems.
\end{abstract}

\section{Introduction}

Accurate measurements of quantum correlations in the next generation of experiments with ultracold atoms in optical lattices are one of the most challenging obstacles in the quest for the realization of quantum simulators of magnetic systems \cite{reviewBloch,reviewAnna}. Not only local order parameters, but also long range correlations are necessary for the faithful discrimination of magnetic phases in the strongly correlated regime. 
One of the advantages of using optical lattices for simulating solid state systems, is that atoms are easily coupled to light with extremely accurate control. The atomic matter properties are then inferred from the measurement of the light scattered off of an atomic sample. 

\begin{figure}[t]
\begin{center}
\includegraphics[scale=0.7]{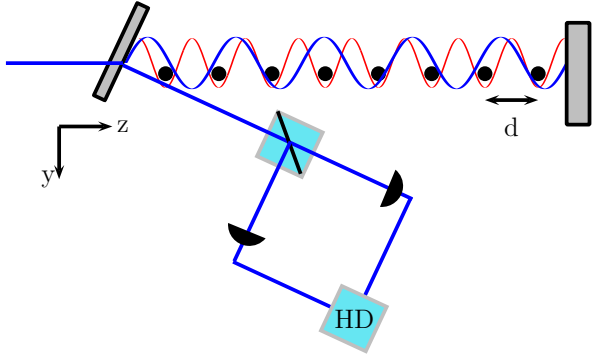}
\caption{(Color online) Schematic detection setup, atoms placed in an optical lattice of periodicity $d/2$ (thin line; red lattice) are illuminated by a laser beam in a standing wave configuration (dark line; blue lattice) shifted by $\alpha$ from the optical lattice configuration. The output light is redirected through a polarimeter which measures its polarization through a homodyne detection (HD).}
\label{fig:setup}
\end{center}
\end{figure}

Here, we review a proposal for inferring spin-spin correlations using a quantum polarization spectroscopy scheme based on a light matter interface\cite{polzik}. The idea, put forward initially in Ref.~\cite{Eckert2008} and further developed in Refs.~\cite{Roscilde2009,dechiara-spectroscopy}, consists in coupling the polarization of a beam illuminating the optical lattice with the magnetic momenta of the trapped atoms. As we explain in this work, using a light beam in a standing wave configuration, as sketched in Fig.~\ref{fig:setup}, one can achieve a modulation of the light-atoms coupling which allows one to reconstruct the atomic spin-spin correlations. One of the most important advantages of this method is that it is non destructive, i.e. the atomic sample is kept in the trap and can be reused for further measurements. We apply this method to a specific one-dimensional spin chain. We show how to measure the model phase diagram by accessing the order parameters of the different phases. 

Furthermore, we will show that apart from measuring spin-spin correlations, polarization spectroscopy allows to discriminate whether a magnetic system, in our case a spin chain, is entangled or not. This is a long standing problem in the context of quantum information theory and many-body systems (see for example \cite{TothReview}), since it is extremely difficult to characterize quantum entanglement for general many particle states. In some specific cases, such as spin systems, one can derive spin squeezing inequalities involving the system total angular momentum which reveals whether the many-body state is entangled or not. In this context, several proposals have been put forward based on particle scattering, e.g. neutron scattering, to probe entanglement in magnetic systems \cite{Wiesniak2005,krammer,cramer}. Here we show that entanglement witnesses based on spin-squeezing inequalities are straightforwardly measured with our proposed scheme. Moreover, the flexibility of optical setups in modulating the periodicity of the probe wave give a lot of freedom and accuracy in the measurements compared to neutron scattering. We expect this scheme to open a new route for the, so far elusive, detection of many-body entanglement.

The paper is organized as follows:
in Sec.~\ref{sec:QPS} we review the quantum polarization spectroscopy technique aimed at measuring spin-spin correlations in optical lattices while in Sec.~\ref{sec:model} we discuss how to simulate spin chains in optical lattices; in Sec.~\ref{sec:results} we show the numerical results for the output signal for the reconstruction of the phase diagram of the spin model; in Sec.~\ref{sec:ent} the entanglement detection using quantum polarization spectroscopy is described, and finally in Sec.~\ref{sec:conclusions} we conclude.

\section{The detection scheme}
\label{sec:QPS}
The detection scheme based on polarization spectroscopy that we use in this work has been described in \cite{Eckert2008,Roscilde2009,dechiara-spectroscopy}. 
The scheme consists in shining the atoms with a non resonant probe beam in a standing wave configuration as shown in Fig.~\ref{fig:setup}. Due to the Faraday effect, the polarization of the incoming light is rotated as a consequence of an effective magnetic field generated by the atomic magnetic moments. By measuring the change in the polarization of the output light we acquire information on the total angular momentum of the atomic sample. 

For light propagating along the z-axis, parallel to the atomic array, light-atom interaction is best expressed using the Stokes parameters defined as:
\begin{eqnarray}
s_1&=&\frac{1}{2}(a^\dagger_x a_x-a^\dagger_y a_y),
\\
s_2&=&\frac{1}{2}(a^\dagger_y a_x+a^\dagger_x a_y),
\\
s_3&=&\frac{1}{2i}(a^\dagger_y a_x-a^\dagger_x a_y),
\end{eqnarray}
where $a_x$ and $a_y$ are the photon annihilation operators with polarization along $x$ and $y$ such that $n_x=a^\dagger_xa_x$ and $n_y=a^\dagger_ya_y$ are the number of photons per unit of time with polarization $x$ and $y$ respectively. Using this definition, the atom-light interaction is described by the Hamiltonian:
\begin{equation}
H_{AL} = -\kappa s_3 J^{eff}_z,
\end{equation}
 where the coupling constant $\kappa$ depends on the optical depth of the atomic sample and on the probability of exciting an atom due to the probe. The effective angular momentum $J^{eff}_z$ depends on the intensity profile of the probe beam. In the case of a simple standing wave it is given by 
\begin{equation}
\label{eq:Jeff}
J^{eff}_z =\frac{1}{\sqrt{L}} \sum_n c_n S_{zn},
\end{equation}
and the coefficients are defined as
 \begin{equation}
c_n = 2 \int dz \cos^2[k_P(z-a)] |w(z-nd)|^2,
\end{equation}
where $k_P$ is the wavevector of the probe light, $a$ is a shift and $w(z-nd)$ is the first band Wannier function of the atom centered at lattice position $z=nd$. In the calculations, for simplicity, we approximate  the Wannier functions with delta functions centered at the lattice positions so that the coefficients are now given by $c_n=2\cos^2[k_Pd(n-\alpha)]$ where we defined the dimensionless shift $\alpha=a/d$.

We assume the incoming light to be strongly polarized along the $x$ direction, i.e. $\langle S_1 \rangle = N_{ph} \gg 1$ where $S_i=\int dt s_i$ and $N_{ph}$ is the beam total number of photons. Therefore we can approximate the other two Stokes operators as two effective conjugated variables: $X=S_2/\sqrt{N_{ph}}$ and $P=S_3/\sqrt{N_{ph}}$ such that 
\begin{equation}
[X,P] = \frac{i S_1}{N_{ph}} \sim i.
\end{equation}
Integrating out the Heisenberg equations of motion for these light quadratures we obtain
\begin{equation}
X_{out} = X_{in} - \kappa J_z^{eff},
\end{equation}
where $X_{in}$ is the quadrature of the incoming light, and $X_{out}$ is the output light emerging from the sample that can be measured using homodyne detection as shown in Fig.~\ref{fig:setup}. Since we assumed the initial beam to be strongly polarized along the $x$ direction, $\langle X_{in}\rangle = 0$, we obtain:
\begin{equation}
\langle X_{out} \rangle = - \kappa \langle J_z^{eff} \rangle,
\end{equation}
thus, the mean of the effective angular momentum is mapped into the mean output light quadrature. Similarly, higher moments (fluctuations) are also mapped.  In this way all the moments of $ J_z^{eff} $ can be extracted by the noise distribution of the output light and in particular the variance:
\begin{equation}
(\Delta X_{out})^2 = (\Delta X_{in})^2 +\kappa^2(\Delta J_z^{eff})^2,
\end{equation}
where $(\Delta X_{in})^2$ is the input noise (for a coherent state ($\Delta X_{in})^2=1/2$).
Different magnetic phases can be distinguished by studying the mean effective angular momentum  $\langle J_z^{eff} \rangle$ and the variance $(\Delta J_z^{eff})^2$. The former immediately tells us whether the spin chain is ferromagnetic or not. The second gives us access to magnetic correlations:
\begin{eqnarray}
\label{eq:eps}
\varepsilon(k_P,\alpha)& =& (\Delta J_z^{eff})^2 =
\nonumber\\
 &=&\frac 4L \sum_{nm}\cos^2[k_P d(m-\alpha)] \cos^2[k_P d(n-\alpha)]
 \nonumber\\
 &\times&\mathcal G_z(m,n).
\end{eqnarray}
where $\mathcal G_z(m,n)\equiv\langle S_{zm} S_{zn}\rangle -\langle S_{zm}\rangle\langle S_{zn}\rangle $ is the two-point correlation function.
As noticed in \cite{Roscilde2009}, in the case of a sample for which the net magnetization is zero, the output signal can be connected to the magnetic structure factor. Indeed by averaging the signal over the phase shift $\alpha$ one gets:
\begin{eqnarray}
\label{eq:eps_structure}
\bar\varepsilon(k_P) \equiv\int d\alpha\varepsilon(k_P,\alpha)=
\frac{1}{2}S(2k_P) 
\end{eqnarray}

In principle, averaging over the phase shift $\alpha$ we are losing some information on the correlations. In this work we assume an accurate control on the shift $\alpha$ and we define the quantity:
\begin{equation}
\label{eq:deltaeps}
\Delta\varepsilon(k_P,\alpha_1,\alpha_2)\equiv \varepsilon(k_P,\alpha_1)-\varepsilon(k_P,\alpha_2)
\end{equation}
which is the difference of the signal with fixed wavevector $k_P$ at two different phase shifts. We will show that the quantity $\Delta\varepsilon(k_P,\alpha_1,\alpha_2)$, by appropriately choosing the parameters $k_P, \alpha_1, \alpha_2$,  can be linked to the local order parameters necessary for the identification of the different phases of a given model.

If we assume no net magnetization, then the expression for  $\Delta\varepsilon$ simplifies to:
\begin{eqnarray}
 \Delta\varepsilon(k_P,\alpha1,\alpha2) &=& \frac{1}{2L}\sum_{mn}\left\{\cos[2k_Pd(m+n-2\alpha_1)] \right .
 \nonumber\\
 &-&\left.\cos[2k_Pd(m+n-2\alpha_2)] \right\} \mathcal G_z(m,n)
\end{eqnarray}

\section{Realization of spin-$1$ Hamiltonians in optical lattices}
\label{sec:model}
Here we briefly review how to simulate spin chains with ultracold atoms in optical lattices and discuss its phase diagram. Spin-1 atoms confined in a deep optical lattice generated by two counter-propagating lasers of wavelength $\lambda$ are well described, within the tight binding approximation, by the Bose-Hubbard Hamiltonian \cite{Jaksch1998}. Defining the creation and annihilation operators $a_{i,\sigma}^\dagger$ and $a_{i,\sigma}$ in site $i$ of an atom with spin components $\sigma=1,0,-1$ along the quantization axis, the Hamiltonian takes the form:
\begin{eqnarray}
H_{BH} &=& \frac{U_0}{2}\sum_i n_i(n_i-1)+\frac{U_2}{2}\sum\left(\bm{S}_i^2-2n_i\right)
-\mu\sum n_i+
\\
&-&t\sum_{i\sigma} \left(a_{i,\sigma}^\dagger a_{i+i,\sigma}+h.c. \right)
\nonumber
\end{eqnarray}
 The operator $n_i = \sum_\sigma a_{i,\sigma}^\dagger a_{i,\sigma}$ is the total number operator of site $i$ while $\bm{S}_i = \sum_{\sigma,\sigma'}a_{i,\sigma}^\dagger \bm{T}_{\sigma,\sigma'} a_{i,\sigma'}$ is the spin operator (matrices $\bm{T}$ are the usual spin-1 angular momentum operators and we use $\hbar=1$).
The parameters appearing in the Hamiltonian $H_{BH}$ are: the usual Hubbard repulsion $U_0$, a spin dependent interaction $U_2$, the chemical potential $\mu$ and the tunneling rate $t$. While the chemical potential fixes the total number of atoms, the remaining parameters can be evaluated from the depth of the optical lattice and from the scattering lengths associated with different scattering channels \cite{Imambekov}.

The phase diagram of this model in the $\mu-t$ plane consists of insulating lobes as in the spinless Bose-Hubbard model where the lobes size depends on the ratio $U_2/U_0$ \cite{Rizzi2005}. For unit filling and for sufficiently small tunneling $t$ the system is in a Mott insulator state with one atom per site. Virtual tunneling of the atoms between neighboring sites gives rise to an effective magnetic interaction described by the bilinear-biquadratic Hamiltonian \cite{Imambekov}:
\begin{eqnarray}
\label{eq:HBB}
H_{BB} 
= J\sum_i \cos(\theta) \bm{S}_i\cdot\bm{S}_{i+1}+ \sin(\theta) (\bm{S}_i\cdot\bm{S}_{i+1})^2
\end{eqnarray}
The Hamiltonian \eqref{eq:HBB} is derived within second order perturbation theory in the ratio $t/U_\alpha$, $\alpha=0,2$ and the relevant parameters read:
\begin{eqnarray}
\tan(\theta) &=& \frac{U_0}{U_0-2U_2},
\\
J &=& \frac{2t^2}{U_0+U_2} \sqrt{1+\tan^2(\theta)},
\end{eqnarray}
where the angle $\theta$ varies in the interval $[-\pi;\pi]$.

\begin{figure}[t]
\begin{center}
\begin{pspicture}(-2.5,-2.5)(2.5,2.5)
\pscircle[linewidth=0.05](0,0){2}
\psline[linewidth=0.05](-1.4142,-1.4142)(1.4142,1.4142)
\psline[linewidth=0.05](0,0)(0,2)
\psline[linewidth=0.05](0,0)(1.4142,-1.4142)
\rput(-1,0){\Large F}
\rput(1,0){\Large H}
\rput(0.5,1.3){\Large C}
\rput(0,-1){\Large D}
\rput(0,2.35){\Large $\frac{\pi}{2}$}
\rput(1.6,1.6){\Large $\frac{\pi}{4}$}
\rput(1.6,-1.6){\Large $-\frac{\pi}{4}$}
\rput(-1.75,-1.7){\Large $-\frac{3\pi}{4}$}
 \end{pspicture} 
\end{center}
\caption{Phase diagram of the bilinear-biquadratic Hamiltonian \eqref{eq:HBB} in the interval $\theta\in[-\pi;\pi]$. The four phases are: the ferromagnetic phase (F), the critical phase (C), the Haldane phase (H) and the dimer phase (D).}
\label{fig:phasediagram}
\end{figure}
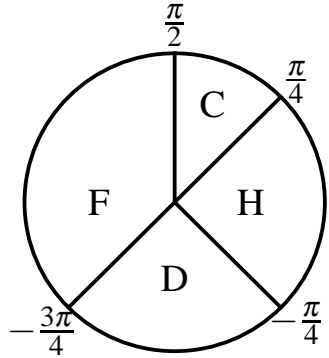

Hamiltonian \eqref{eq:HBB} is characterized by a rich phase diagram, sketched in Fig.~\ref{fig:phasediagram}, depending on the angle $\theta$ and which has  been extensively studied in the literature, see \cite{AKLT,Chubukov,FathSolyom1991,FathSolyom1995, Schollwock1996,Buchta2005, Rizzi2005,Lauchli2006,Reed,Schollwock_book} and references therein. Here we briefly discuss the model phase diagram and the corresponding order parameters.

\emph{The ferromagnetic phase.-} For $\pi/2 <\theta < 5\pi/4$ the ground state is ferromagnetic: all the spins, breaking the rotational symmetry of $H_{BB}$, align along some direction with a net spontaneous magnetization, which serves as a local order parameter. For the remaining values of $\theta$ the ground state lacks of spontaneous magnetization. However, within this interval we can distinguish different phases.

\emph{The critical phase.-} In the interval $\pi/4<\theta<\pi/2$ the system is in a critical phase in which the model is gapless due to soft collective modes at momenta $q = 0,\pm 2\pi/(3d)$ where $d=\lambda/2$ is the distance between two adjacent sites. The ground state organizes in slightly correlated clusters of three neighboring spins (trimers). This fact is reflected in the spin-spin correlation functions $\langle S_{zi}S_{z(i+r)}\rangle $ which show a period-$3$ oscillations \cite{FathSolyom1991}. In momentum space this feature emerges as a peak at $q= 2\pi/(3d)$ in the magnetic structure factor defined as:
\begin{equation}
S(q) = \frac 1L \sum_{mn} e^{i qd(m-n)} \langle S_{zm}S_{zn}\rangle.
\end{equation}
Recently L\"auchli et al. \cite{Lauchli2006} have shown that nematic (i.e. quadrupolar) correlations at momentum $q= 2\pi/(3d)$ are enhanced in the critical phase while spin correlations become smaller when increasing $\theta$ from $0.2\pi$ to $0.5 \pi$.  Together with the absence of the gap, the enhanced nematic correlations are a distinctive feature of the critical phase. 
 
\emph{The Haldane phase.-}  For $-\pi/4<\theta<\pi/4$ the system is in the Haldane phase which is gapped and contains for $\theta=0$ the spin-1 isotropic Heisenberg chain and for $\tan(\theta) = 1/3$ the Affleck-Kennedy-Lieb-Tasaki (AKLT) point for which the ground state is exactly known \cite{AKLT}. Numerical results  in this region based on density matrix renormalization group (DMRG) simulations show that decreasing $\theta$ from $\pi/4$ to $\theta_L \simeq 0.1314 \pi$, the so called Lifshitz point, the peak at momentum $q= 2\pi/(3d)$ in the magnetic structure factor moves continuously to $q=\pi/d$ (see Ref.~\cite{Schollwock1996}). Notice that, although these peaks signal some correlations, the presence of a gap excludes local long range magnetic order and spin correlations decay exponentially. The Haldane phase can be instead characterized in terms of a hidden topological order parameter, called the string order parameter \cite{Rommelse}:
 \begin{equation}
O_\pi(m,n) =\left \langle S_{zm} \exp\left(i \pi \sum_{l=m-1}^{n-1} S_{zl}\right) S_{zn}\right\rangle
\end{equation}
This order, being topological, cannot be revealed with local measurements.

\emph{The dimer phase.-}   The interval $-3\pi/4<\theta<-\pi/4$ is still debated. At $\theta=-\pi/4$ the gap closes and for smaller values of $\theta$ it reopens again. In this region the ground state breaks translational invariance and organizes in slightly correlated dimers. For $-3\pi/4<\theta<-\pi/2$ it is still under debate whether the system is always dimerized or it becomes  nematic as proposed by Chubukov \cite{Chubukov}. Numerical results \cite{FathSolyom1995,Buchta2005,Rizzi2005,Lauchli2006} show that the dimer order parameter:   
\begin{equation}
\label{eq:dimer_order_parameter}
D= |\langle H_i-H_{i+1}\rangle|
\end{equation}
where $H_i=\cos(\theta) \bm{S}_i\cdot\bm{S}_{i+1}+ \sin(\theta) (\bm{S}_i\cdot\bm{S}_{i+1})^2$,
is different from zero up to values very close to $\theta=-3\pi/4$ giving strong evidence for the absence of the nematic phase except only in an infinitesimal region close to $\theta=-3\pi/4$.

\section{Phase diagram reconstruction}
\label{sec:results}
\begin{figure}[t!]
\begin{center}
\includegraphics[scale=0.8]{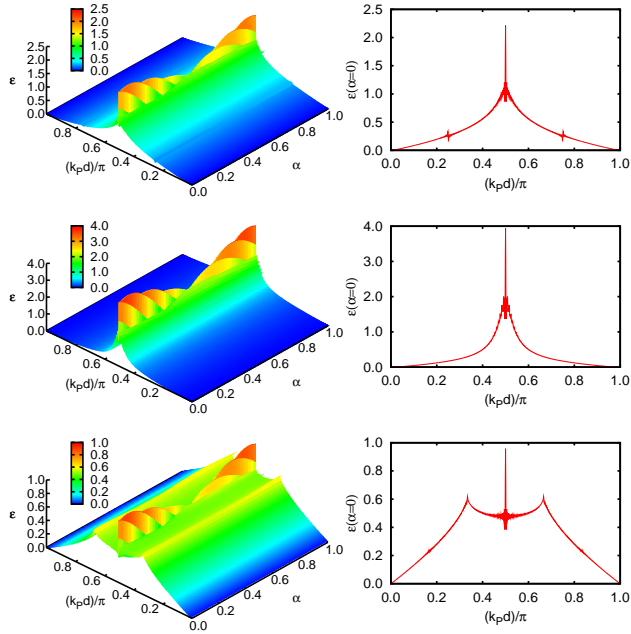}
\caption{Left column, the function $\varepsilon(k_P,\alpha)$ for different values of $\theta$ in the three phases for $L=132$: top $\theta = -0.5\pi$ (dimer), middle $\theta=0$ (Haldane), bottom $\theta=0.3\pi$ (critical). Right column, the same plots but restricted to $\alpha=0$.}
\label{fig:epsall}
\end{center}
\end{figure}

In this section we discuss the results of the detection scheme applied to the bilinear-biquadratic Hamiltonian. The quantities $\varepsilon(k_P,\alpha)$, which depend on all possible correlations between two spins, are computed numerically by means of the DMRG algorithm \cite{dmrg}. We simulate spin chains with open boundary conditions and lengths which are multiple of $2$ and $3$ reducing known finite size effects due to incommensurability \cite{FathSolyom1991}. In the DMRG simulations we choose the number of block states sufficiently large to ensure that the truncation error is less than $10^{-6}$.

The ferromagnetic phases is easily detected by looking at the average value of the effective angular momentum:
\begin{equation}
\langle J_z^{eff}(k=0)\rangle =\frac{2}{\sqrt{L}} \sum_n S_{zn} 
\end{equation}
which is proportional to the total magnetization along the $z$ direction.

Since $\langle J_z^{eff}\rangle$ is zero in the other three phases, we need the second moment of $J_z^{eff}$ in order to characterize this phase. In Fig.~\ref{fig:epsall} we show  $\varepsilon(k_P,\alpha)$ in the critical, Haldane and dimer phases. A common feature of the three phases is the presence of a high peak at $k_P d=\pi/2$ due to antiferromagnetic correlations. Apart from this, the three plots are qualitative different. In fact for $\theta>\theta_L$, the Lifshitz point, the signal is characterized by peaks at $k_Pd\sim\pi/3$ and $k_Pd\sim2\pi/3$. These resemble the peaks of the magnetic structure factor\footnote{Notice that from Eq.~\eqref{eq:eps_structure}, $\varepsilon(k_P,\alpha)$ is related to the structure factor $S(2k_P)$ at double the value of the momentum.} and are due to the period-3 oscillations of the correlation functions. We will study these correlations in Sec.~\ref{sec:critical} and show that they detect the critical phase. For $\theta < -\pi/4$ we find the appearance of other small peaks at $k_P d = \pi/4$ and $k_P d = 3\pi/4$ signaling a different order with a larger period. We will study more carefully these features in Sec.~\ref{sec:dimer}.

Since the presence of these distinctive peaks is relevant for the determination of the phase of the spin chain, we find it convenient to subtract the background generated by all possible correlations in definition \eqref{eq:eps} by instead using the quantity $\Delta\varepsilon(k_P,\alpha_1,\alpha_2)$ defined in Eq.~\eqref{eq:deltaeps}.

\begin{figure}[t]
\begin{center}
\includegraphics[scale=0.3]{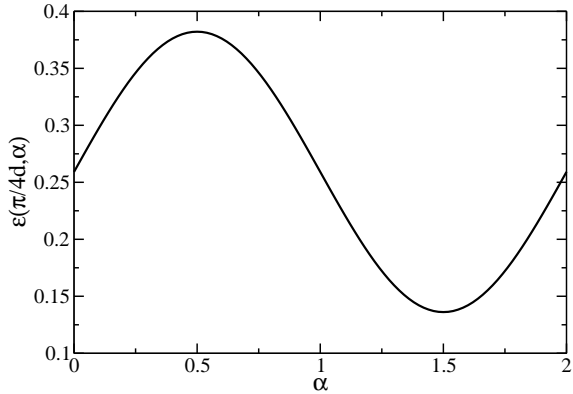}
\caption{The quantity $\varepsilon(\pi/4d,\alpha)$ for $\theta=-0.5 \pi$ (dimer phase)  for $L=132$ as a function of $\alpha$.}
\label{fig:epscritk14}
\end{center}
\end{figure}

To see how to choose the parameters $k_P,\alpha_1,\alpha_2$, let us consider the dimer phase. In this case we find it convenient to choose $k_P=\pi/4d$ which is the periodicity of the dimers. Then we study the behavior of $\varepsilon(\pi/4d,\alpha)$ in one point of the dimer phase as a function of $\alpha$ as shown in Fig.~\ref{fig:epscritk14}. The quantity $\varepsilon(\pi/4d,\alpha)$ is an oscillating function of $\alpha$. In order to optimize the information on the correlations at $k_P=\pi/4d$ we choose the difference between the maximum at $\alpha_1=3/2$ and the minimum at $\alpha_2=1/2$. Thus as an indicator of the critical phase, we will study the quantity $\Delta\varepsilon(\pi/4d,3/2,1/2)$. In the critical phase, a similar analysis leads to $k_P=\pi/4d, \alpha_1=5/4,\alpha_2=1/2$ (see also \cite{dechiara-spectroscopy}).

\subsection{Detecting the critical phase}
\label{sec:critical}
For the critical phase, we have seen that the distinctive peaks are at $k_P d=\pi/3$, and as shown in the previous section we choose $\alpha_1=5/4$ and $\alpha_2=1/2$. Thus we define the quantity  \begin{eqnarray}
 C_\varepsilon&=&\Delta\varepsilon(\pi/3d,5/4,1/2) =
\nonumber\\
&=& \frac{1}{L} \sum_{mn} \cos\left[\frac{2\pi}{3} (m+n) +\frac{\pi}{3}\right] \mathcal G_z(m,n),
\end{eqnarray}
where we used the fact that the ground state is an eigenstate of the total angular momentum with zero eigenvalue:
\begin{equation}
\sum_n \langle S_{zm}S_{zn}\rangle = \langle S_{zm}\sum_n S_{zn}\rangle =0
\end{equation}

\begin{figure}[t]
\begin{center}
\includegraphics[scale=0.7]{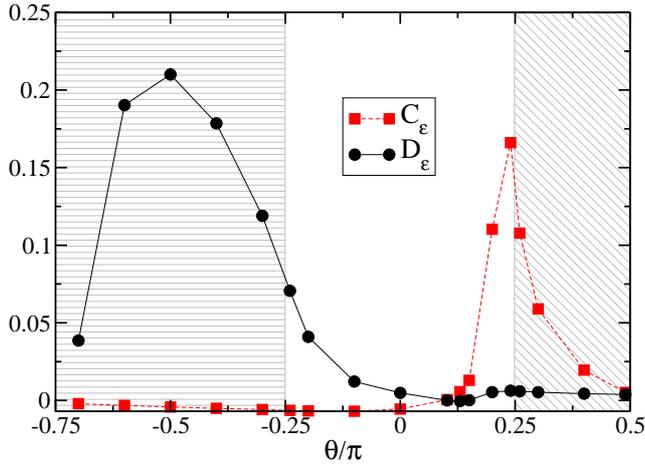}
\caption{(Color online) The quantities $C_\varepsilon=\Delta\varepsilon(\pi/3d,5/4,1/2)$ (squares) and $D_\varepsilon =\Delta\varepsilon(\pi/4d,1/2,3/2)$ (circles) as a function of $\theta$  for $L=132$.  We distinguish the model phases with different shading: horizontal lines (dimer), no shading (Haldane), oblique lines (critical). The solid and dashed lines are only guides to the eye.}
\label{fig:criticaldimer}
\end{center}
\end{figure}

The quantity $ C_\varepsilon$ is sensitive to correlations which oscillate with a period 3 and represents a footprint of the critical phase. In fact, in Fig.~\ref{fig:criticaldimer} we show the signal $ C_\varepsilon$ for different values of $\theta$ in the antiferromagnetic phase between $-0.7\pi$ and $0.5\pi$. The results clearly show that the critical phase is very well detected by a positive value of  $ C_\varepsilon$. For $\theta=0.2 \pi$, in the Haldane phase and close to the phase transition, we still observe a large positive value, probably due to residual period 3 correlations persisting in the Haldane phase for $\theta>\theta_L$. However for $\theta=0.2\pi$ we find a non negligible dependence with the size of the sample. A finite size scaling analysis suggests that in the thermodynamical limit for $L\to\infty$ the quantity $ C_\varepsilon$ goes to zero as $1/L$ for $\theta=0.2\pi$, while for the other values of $\theta \ge 0.24 \pi$ it converges to a finite value (see Ref.~\cite{dechiara-spectroscopy}).
Our findings indicate that by measuring $ C_\varepsilon$ which depends only on spin-spin interactions we are able to infer the occurrence of the phase transition and thus the quantity $C_\varepsilon$ behaves as an order parameter for the critical phase.

\subsection{Detecting the dimerized phase}
\label{sec:dimer}
Let us now consider the dimerized phase. As discussed before the presence of peaks at $k_P=\pi/4d$ signals  pairing of neighboring spins. Notice that if we average the signal $\varepsilon(k_P,\alpha)$ over $\alpha$ these peaks disappear. Therefore these features are not visible in the magnetic structure factor.

We find that the quantity 
\begin{eqnarray}
 D_\varepsilon &\equiv&\Delta\varepsilon(\pi/4d,1/2,3/2)
\nonumber\\
&=& -\frac 1L \sum_{mn} \sin\left[\frac{\pi}{2} (m+n)\right] \mathcal G_z(m,n)\end{eqnarray}
 is suitable for the detection of the dimer phase. The factor $\sin\left[\pi/2 (m+n)\right]$ ensures that only the pairs of spins with positions $m$ and $n$ of opposite parity contribute to $ D_\varepsilon$. Moreover the $\sin$ function gives an alternating sign depending on whether the distance between the sites is even or odd. Therefore the quantity $ D_\varepsilon$ is an extension to long range correlations of the dimer order parameter $D$ defined in Eq.~\eqref{eq:dimer_order_parameter}.

In Fig.~\ref{fig:criticaldimer} we show the results for the signal $ D_\varepsilon$ for different values of $\theta$. Similar to the dimer order parameter $D$, the quantity $ D_\varepsilon$ is significantly different from zero only in the dimerized phase, therefore acting as an alternative dimer order parameter.

\section{Entanglement detection}
\label{sec:ent}
Detecting entanglement in many-body systems is not an easy task. In magnetic systems, such as the spin chain considered in this work, one can employ spin squeezing inequalities based on collective angular momentum operators (see \cite{TothReview} for a review). An entanglement witness is an operator which is positive valued for all separable (non entangled) states, while there exists at least one entangled state for which the expectation value of the witness is negative.
The witness we  propose is based on the effective angular momentum defined in \eqref{eq:Jeff}. The construction follows Refs.~\cite{TothReview,Wiesniak2005}.

As before we define an effective angular momentum which we assume can be  written on the light fluctuations:
\begin{equation}
J_\alpha = \sum_m c_m S_{\alpha m} \qquad \alpha=x,y,z
\end{equation}
where now we consider the angular momentum fluctuations in the two other directions.
Let us consider the quantity:
\begin{eqnarray}
V &=& \sum_{\alpha=x,y,z} \Delta J_\alpha^2 =  
\sum_{\alpha=x,y,z}\sum_{ij} c_i c_j (\langle S_{\alpha i}S_{\alpha j}\rangle- \langle S_{\alpha i}\rangle\langle S_{\alpha j}\rangle)
\end{eqnarray}
Now if the many-body system is in a product state:
\begin{equation}
\rho_{prod} = \rho_1 \otimes \rho_2 \otimes\dots\otimes \rho_N
\end{equation}
we have:
\begin{equation}
\langle S_{\alpha i}S_{\alpha j}\rangle- \langle S_{\alpha i}\rangle\langle S_{\alpha j}\rangle = \delta_{ij}\left( \langle S_{\alpha i}^2\rangle- \langle S_{\alpha i}\rangle^2\right)
\end{equation}
Using the relation for spin $s$ particles:
\begin{equation}
\langle S_{xi}^2\rangle+\langle S_{yi}^2\rangle+\langle S_{zi}^2\rangle = s(s+1)
\end{equation}
and the inequality:
\begin{equation}
\langle S_{xi}\rangle^2+\langle S_{yi}\rangle^2+\langle S_{zi}\rangle^2 \le s^2
\end{equation}
we see that for product states:
\begin{equation}
\label{eq:prod}
V_{prod}\ge s\sum_i c_i^2
\end{equation}
If we consider separable states:
\begin{equation}
\rho_{sep} = \sum_n p_n \rho_{n,sep}, \quad 0<p_n<1, \quad \sum_n p_n=1
\end{equation}
where each state $\rho_{n,sep}$ in the mixture is separable, we have
\begin{equation}
V_{sep}=\sum_\alpha \Delta J_\alpha^2 \ge \sum_n p_n \sum_\alpha (\Delta J_\alpha^2)_n 
\ge \sum_n p_n s\sum_i c_i^2 = s\sum_i c_i^2 
\end{equation}
where the first inequality comes for a mixture $\rho=\sum_n p_n \rho_n$: $\Delta X^2 \ge \sum_n p_n (\Delta X^2)_n$ and $(\Delta X^2)_n$ is the variance evaluated in the $n$th ensemble element; the second inequality comes from Eq.~\eqref{eq:prod}.

\begin{figure}[t]
\begin{center}
\includegraphics[scale=0.7]{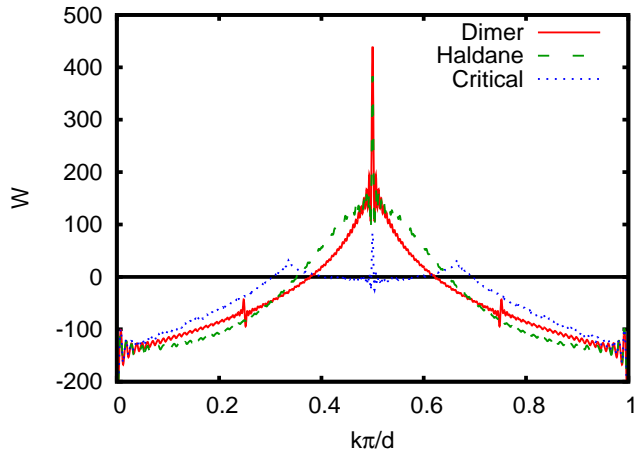}
\caption{(Color Online) Expectation value of the entanglement witness $W$ from Eq.~\eqref{eq:w1} for three values of $\theta$ in the three different phases: $\theta=-0.5\pi$ (solid (red), dimer phase) $\theta= 0.102 \pi$ (dashed (green), AKLT point in the Haldane phase), $\theta= 0.3\pi$ (dotted(blue), critical phase). In the numerical simulations we take $L=96$. All these states are clearly detected for small enough values of $k$.}
\label{fig:wit1}
\end{center}
\end{figure}

Therefore, a possible entanglement witness is given by the quantity:
\begin{equation}
\label{eq:w1}
W = V-s\sum_i c_i^2
\end{equation}
Notice that the coefficients $c_i$ and consequently the quantity $V$ depend on the probe light momentum $k$ and on the shift, $a$, between the optical lattice and the probe light on the standing wave configuration. Both parameters can be changed giving,  therefore, an important and necessary flexibility for detection of different entangled states.
In Fig.~\ref{fig:wit1} we show $W$ for $a=0$ as a function of $k$ for states in the critical, Haldane and dimer, phases. It is evident that for certain values of $k$, $W$ is negative thus detecting entanglement.

This method provides an operational entanglement detection scheme which is scalable, robust and that can be realized in present-day experiments with ultracold atoms in optical lattices. We stress that the  quantity $W$ is very general and can be used even if the sample is subject to thermal fluctuations or disorder.

\section{Conclusions}
\label{sec:conclusions}
We have presented a probing technique based on matter-light interface for the investigation of quantum correlations in magnetic systems simulated by ultracold atoms in optical lattices. We have shown that this scheme permits to obtain experimentally  the order parameters of non trivial magnetic phases by homodyne measuring the fluctuations of the probing light quadratures after crossing the atomic sample. Moreover, we have shown that this technique, which is not destructive and realizable with present technology, allows also to detect experimentally the entanglement in magnetic non trivial many body systems without actually carrying out an unnecessary state tomography.

\begin{acknowledgements}
We thank Oriol Romero-Isart for frutiful discussions. We acknowledge support from the Spanish MICINN (Juan de la Cierva, FIS2008-01236 and QOIT-Consolider Ingenio 2010), Generalitat de Catalunya Grant No. 2005SGR-00343. We used the DMRG code available at {\tt http://www.dmrg.it}.
\end{acknowledgements}

\pagebreak


\begin{thebibliography}{99}

\bibitem{reviewBloch}
I. Bloch, J. Dalibard, and W. Zwerger, Rev. Mod. Phys. {\bf 80}, 885 (2008).
\bibitem{bloch-qpt} 
M. Greiner, O. Mandel, T. Esslinger, T. W. H\"{a}nsch, and I. Bloch, Nature {\bf 415}, 39 (2002).

\bibitem{reviewAnna}
M. Lewenstein, A. Sanpera, V. Ahufinger, B. Damski, A. Sen De, U. Sen,  Adv. in Phys. {\bf 56},243 (2007). 

\bibitem{polzik}
J. L. S\o rensen, J. Hald, and E. S. Polzik,
Phys. Rev. Lett. {\bf 80}, 3487 (1998);
J. L. S\o rensen, J. Hald, N. J\o rgensen, and E. S. Polzik, J. Mod. Opt. {\bf 44}, 1917 (1997).


\bibitem{Eckert2008}
K. Eckert, O. Romero-Isart, M. Rodriguez, M. Lewenstein, E. S. Polzik, and A. Sanpera, Nat. Phys. {\bf 4}, 50 (2008). 

\bibitem{Roscilde2009}
T. Roscilde, M. Rodriguez, K. Eckert, O. Romero-Isart, M. Lewenstein, E. Polzik, A. Sanpera,
New. J. Phys. {\bf 11}, 055041 (2009).

\bibitem{dechiara-spectroscopy}
G. De Chiara, O. Romero-Isart, and A. Sanpera,  
Phys. Rev. A {\bf 83}, 021604(R) (2011).

\bibitem{TothReview}
O. G\"uhne and G. T\'oth, Phys. Rep. {\bf 474}, 1 (2009).



\bibitem{Wiesniak2005}
M. Wiesniak, V. Vedral, and C. Brukner, New. J. Phys. {\bf 7}, 258 (2005).

\bibitem{krammer}
P. Krammer, H. Kampermann, D. Bru\ss, R. A. Bertlmann, L. C. Kwek, and C. Macchiavello,
Phys. Rev. Lett. 103, 100502 (2009).

\bibitem{cramer}
M. Cramer, M. B. Plenio, and H. Wunderlich,
Phys. Rev. Lett. {\bf 106}, 020401 (2011).


\bibitem{dmrg}
S. R. White, Phys. Rev. Lett. {\bf 69}, 2863 (1992);
U. Schollw\"ock, Rev. Mod. Phys. {\bf 77}, 259 (2005);
G. De~Chiara, M. Rizzi, D. Rossini, and S. Montangero, J. Comp. Theor. Nanos. {\bf 5}, 1277 (2008).

\bibitem{Jaksch1998}
D. Jaksch, C. Bruder, J. I. Cirac, C. W. Gardiner, and P. Zoller,
Phys. Rev. Lett. {\bf 81}, 3108 (1998).

\bibitem{Imambekov}
A. Imambekov, M. Lukin, and E. Demler, 
Phys. Rev. A {\bf 68}, 063602 (2003).

\bibitem{Rizzi2005}
M . Rizzi, D. Rossini, G. De Chiara, S. Montangero, and R. Fazio,
Phys. Rev. Lett. {\bf 95}, 250404 (2005).



\bibitem{AKLT}
I. Affleck, T. Kennedy, E. H. Lieb, H. Tasaki, Phys. Rev. Lett. {\bf 59}, 799 (1987).

\bibitem{Chubukov}
A. V. Chubukov, Phys. Rev. B {\bf 43}, 3337 (1991).


\bibitem{FathSolyom1995}
G. F\'ath and J. S\'olyom, Phys. Rev. B {\bf 51}, 3620 (1995).

\bibitem{Schollwock1996}
U. Schollw\"ock, Th. Jolicoeur, and T. Garel,
Phys. Rev. B {\bf 53}, 3304 (1996).


\bibitem{Buchta2005}
K. Buchta, G. F\'ath, \"O. Legeza, and J. S\'olyom, Phys. Rev. B {\bf 72}, 054433 (2005).



\bibitem{Lauchli2006}
A. L\"auchli, G. Schmid, and S. Trebst,
Phys. Rev. B {\bf 74}, 144426 (2006).

\bibitem{Reed}
 P. Reed, J. Phys. A {\bf 27}, L69 (1994). 


\bibitem{Schollwock_book}
U. Schollw\"ock, J. Richter, D. Farnell, and R. Bishop, ``Quantum Magnetism'', Springer-Verlag and Berlin (2004). 

\bibitem{FathSolyom1991}
G. F\'ath and J. S\'olyom, Phys. Rev. B {\bf 44}, 11836 (1991).


\bibitem{Rommelse}
Marcel den Nijs and Koos Rommelse,
Phys. Rev. B {\bf 40}, 4709 (1989).




\end{thebibliography}
\end{document}